\pdfoutput=1 
\documentclass[aps,prd,floatfix,nofootinbib,superscriptaddress,twocolumn]{revtex4-1}

\usepackage{graphicx}
\usepackage{amsmath}
\usepackage{courier} 
\usepackage{longtable}

\usepackage[colorlinks,pdfstartview=FitH]{hyperref}
\hypersetup{linkcolor=blue,citecolor=blue,filecolor=black,urlcolor=blue}

\graphicspath{{Figures/}}

\newcommand{\commentout}[1]{{}}

\newcommand{\pd}[2]{{\frac{\partial #1}{\partial #2}}}

\newcommand{\pdconst}[3]{{\left(\frac{\partial #1}{\partial #2}\right)_{#3}}}

\def\nbar{\bar{n}}

\begin{document}


\title[]{Conserved Charge Susceptibilities in a Chemically Frozen Hadronic Gas}
\author{Jackson Ang'ong'a}
\email[e-mail: ]{angonga2@illinois.edu}
\affiliation{
Department of Physics and Astronomy,
Colgate University, Hamilton, NY 13346
}
\affiliation{
Department of Physics,
University of Illinois at Urbana-Champaign,
Urbana, IL, 61801
}

\author{Todd Springer}
\email[e-mail: ]{gspringer@colgate.edu}
\affiliation{
Department of Physics and Astronomy,
Colgate University, Hamilton, NY 13346
}


\date{August 10, 2015}

\begin{abstract}
  In a hadronic gas with three conserved charges (electric charge,
  baryon number, and strangeness) we employ the hadron resonance gas
  model to compute both diagonal and off-diagonal susceptibilities.
  We model the effect of chemical freeze-out in two ways: one in which
  all particle numbers are conserved below the chemical freeze-out
  temperature and one which takes into account resonance decays.  We
  then briefly discuss possible implications these results may have on
  two active areas of research, hydrodynamic fluctuations and the
  search for the QCD critical point.
\end{abstract}

\keywords{}
\maketitle

\section{Introduction}
Analysis of fluctuations has been of great interest to the heavy ion
collision physics community in recent years.  One line of research
examines the impact of pre-equilibrium initial state fluctuations on
observed final state momentum anisotropies (see \cite{Gale:2013da} for
a review).  In addition to initial state effects, fluctuations of a
thermal nature will occur after the system formed in a heavy ion
collision has attained local equilibrium.  The enhancement of such
fluctuations is an expected experimental signature of the QCD critical
point \cite{Stephanov:1998dy, Stephanov:1999zu, Hatta:2003wn}.  The
principle goal of the Beam Energy Scan at RHIC is the discovery of the
QCD critical point, and the STAR collaboration has recently reported
results in this direction \cite{Adamczyk:2013dal}.  

Thermal fluctuations can also lead to observable particle correlations
\cite{Kapusta:2011gt, Springer:2012iz, Ling:2013ksb}.  These
correlations can be extended over a large pseudorapidity range because
they are sourced by fluctuations which propagate or diffuse according
to the equations of hydrodynamics.  Such correlations are
experimentally studied by measuring balance functions
\cite{Abelev:2013csa, Aggarwal:2010ya}.

To examine the effect of (thermal) conserved charge fluctuations
over the history of a heavy-ion collision, one needs 
thermodynamical input, as the magnitude of such fluctuations are
controlled by the susceptibilities
\begin{eqnarray}
  \chi_{\alpha \beta} \equiv \pd{^2P}{\mu_\alpha \partial \mu_\beta}
\end{eqnarray}
where $P$ is the pressure, $\mu$ denotes chemical potential, and the
abstract indices $\alpha$, $\beta$ denote the particular conserved
charge under consideration.  In this work, $\alpha$ can take on the
value $Q, B, S$ for electric charge, baryon number, or strangeness.
(In fact, as explained in \cite{Ling:2013ksb}, the relevant quantity
is actually $\chi_{\alpha \beta} T / s$ where $s$ is the entropy
density.)  One obvious source of input on these thermodynamic
quantities is lattice QCD, as the susceptibilities have been
calculated there \cite{Borsanyi:2010cj, Borsanyi:2011sw,
  Bazavov:2012jq}.  However, as the lattice assumes full thermal and
chemical equilibrium, its results may not accurately reflect the
thermodynamics in a heavy ion collision where chemical equilibrium is
not always maintained.  It is the effect of chemical freeze-out on the
conserved charge susceptibilities which we wish to explore in this
work.  For the highest energy collisions at RHIC and LHC, the net
charge, baryon number, and strangeness of the fireball are
approximately zero.  Hence, for the remainder of this paper we
calculate susceptibilities at $\mu_B = \mu_Q = \mu_S = 0$.

The effect of chemical freeze-out on thermodynamics in heavy ion
collisions has a long history going back more than two decades
\cite{Bebie:1991ij, Hirano:2002ds, Teaney:2002aj, Huovinen:2007xh,
  Huovinen:2009yb}.  However, these discussions have exclusively
focused on pressure $P$ and energy density $\varepsilon$, with
emphasis the equation of state $P(\varepsilon)$ as this is necessary
input for hydrodynamic simulations.  It was found that despite the
fact that both $P(T)$ and $\varepsilon(T)$ are both modified after
including the effects of chemical freeze-out, $P(\varepsilon)$ is
affected only slightly \cite{Teaney:2002aj}.

Given the recent interest in fluctuations we thus wish to reopen the
line of inquiry with particular attention paid to the
susceptibilities.  To accomplish this, we employ the hadron resonance
gas model (HRG) \cite{Dashen:1969ep, Venugopalan:1992hy,
  Becattini:2004td} which (in full equilibrium) has proved to be an
excellent approximation to lattice calculations at low temperatures
($T\lesssim 180$ MeV).  By including the effects of chemical
non-equilibration in the HRG, we expect the results to more accurately
represent the thermodynamics of the system formed in a heavy ion
collision.

The following is the outline of the paper.  In
Sec. \ref{HRGthermoSec}, we review the thermodynamics of the HRG
model.  We then proceed to detail two ways of modeling chemical
freeze-out.  We refer to the first as ``Full chemical Freeze Out''
(FFO) and explain the details in Sec. \ref{FFOSec} wherein all number
changing processes cease below the freeze-out temperature $T < T_{\rm
  ch}$ and hence all particle numbers are constant.  We also provide
some new analytical formulas for the chemical potentials and
susceptibilities.  In Sec. \ref{PCESec}, we detail the second model of chemical freeze-out
referred to as Partial Chemical Equilibrium (PCE) \cite{Bebie:1991ij}
which permits resonances with short lifetimes to decay.  We present the
results for the susceptibilities in Sec. \ref{ResultsSec}.  We comment
on interesting features of the results, possible phenomenological
implications and directions for future work in
Sec. \ref{ConclusionSec}.

\section{Review of Hadron Resonance Gas Thermodynamics}
\label{HRGthermoSec}
At temperatures below the deconfinement temperature ($T \lesssim 180\,
{\rm MeV}$), the thermalized matter left in the wake of a heavy ion
collision can be modeled as a gas of non-interacting, point-like
hadrons.  The effect of interactions is incorporated by including
resonances.  This model is parameter-free, and does a remarkably good
job of approximating lattice QCD data at sufficiently low temperatures
\cite{Bazavov:2012jq}.

\subsection{Pressure, Entropy Density, Number Density}
To access relevant thermodynamic quantities, one may start by considering the
partial pressure of a given particle species (labeled with the subscript `i').  Assuming
the momentum distribution is isotropic, the expression is
\begin{eqnarray}
  P_i(T,\mu_i) = \frac{g_i}{6 \pi^2}\int_0^\infty \frac{p^4}{E_i}\left[\frac{dp}{e^{(E_i - \mu_i)/T} \pm 1}\right]
\label{partialpressure}
\end{eqnarray}
where the upper/lower signs refer to fermions/bosons, $E_i^2 = p^2 +
m_i^2$ and $g_i$ is a spin degeneracy factor, We have included a
chemical potential $\mu_i$ associated with this particular particle.
It is more convenient to write $\mu_i$ in
terms of chemical potentials associated with the conserved charges:
baryon number (B), electric charge (Q), and strangeness (S) as we will see shortly.

The number density of this particle is found by
\begin{eqnarray}
  n_i(T, \mu_i) = \pdconst{P}{\mu_i}{T} = \frac{g_i}{2 \pi^2} \int_0^\infty \frac{p^2 dp}{e^{(E_i - \mu_i)/T} \pm 1}
\end{eqnarray}
The second equality is a more familiar expression for the number
density, and can be found by first taking the derivative of
(\ref{partialpressure}) and then performing an integration by parts.

We will also need an expression for the partial entropy
\begin{eqnarray}
  s_i(T,\mu_i) &=& \pdconst{P}{T}{\mu_i}\\
  &=& \frac{g_i}{2 \pi^2 T}\int_0^\infty \left[E_i - \mu_i + \frac{p^2}{3E_i} \right]\frac{p^2dp}{e^{(E_i-\mu_i)/T} \pm 1}\nonumber 
\end{eqnarray}
The second equality can be found by first taking the derivative of (\ref{partialpressure}) and then performing integration by parts, or more directly by using the thermodynamic identity
\begin{eqnarray}
  T s_i = \varepsilon_i + P_i - \mu_i n_i
\end{eqnarray}
where $\varepsilon_i$ is the energy density of the particle ``i''.  In
order to compute the \emph{total} entropy density $s$ and/or the total
pressure $P$, one must sum over all hadronic species $i$,
\begin{eqnarray}
  P(T,\{\mu\})  = \sum_i P_i(T,\mu_i).
\end{eqnarray}
Note that the total pressure is a function of the set of all chemical
potentials, which we denote $\{\mu\}$.  The HRG is parameter free, but
one must choose which particles to include in a given calculation.  We
include particles and resonances listed by the particle data group
(PDG) \cite{Agashe:2014kda} with masses less than or equal to 2 GeV. 
More details on our included particles and their decays can be found in Appendix \ref{PDGappendix}.

\subsection{Conserved Charges and Susceptibilities in Full Equilibrium (FE)}
We are interested in fluctuations of conserved charges; as such it
pays to work with chemical potentials associated with these conserved
charges only.  In later sections, when we consider chemical
freeze-out, certain particle numbers are conserved and hence we will
introduce additional conserved ``charges'' and chemical potentials.

The total density of charge $\alpha$ in the system is
\begin{eqnarray}
  n_\alpha(T, \{\mu\}) = \sum_i \alpha_i n_i(T,\mu_i)
  \label{chempot1}
\end{eqnarray}
where $\alpha_i$ is the conserved charge $\alpha$ of the $i$th
particle.  For example, the electric charge density can be
found by setting the abstract index $\alpha = Q$:
\begin{eqnarray}
  n_Q(T, \{\mu\}) = \sum_i Q_i n_i(T, \mu_i) 
\end{eqnarray}

Every conserved charge has a corresponding chemical potential,
so by definition we could alternatively write
\begin{equation}
  n_\alpha(T,\{\mu\}) = \sum_i \pdconst{P_i(T,\mu_i)}{\mu_\alpha}{T} = \sum_i n_i(T,\mu_i) \pdconst{\mu_i}{\mu_\alpha}{T}.
  \label{chempot2}
\end{equation}
Comparing (\ref{chempot1}) with (\ref{chempot2}) we see that
\begin{eqnarray}
  \alpha_i =  \pdconst{\mu_i}{\mu_\alpha}{T}.
\end{eqnarray}
This can only be satisfied if the ``particle'' chemical potentials are related to the
conserved charge chemical potentials as
\begin{eqnarray}
  \mu_i = \sum_\alpha \alpha_i \mu_\alpha
\end{eqnarray}
or, in less abstract notation,
\begin{eqnarray}
  \mu_i = Q_i \mu_Q + B_i \mu_B + S_i \mu_S
  \label{muiFFO}
\end{eqnarray}
where $Q_i, B_i, S_i$ are the electric charge, baryon number, and
strangeness of the $i$th particle.

We now consider the (3 $\times$ 3) matrix of susceptibilities
\begin{eqnarray}
  \chi_{\alpha \beta} &=& \pd{n_\alpha(T,\{\mu\}) }{\mu_\beta} 
  =\sum_i \alpha_i \beta_i \pd{n_i(T,\mu_i)}{\mu_i} 
\end{eqnarray}
In terms of thermodynamic integrals, the components of the susceptibility matrix are
\begin{equation}
  \chi_{\alpha \beta}(T,\{\mu \}) = \sum_i \frac{g_i \alpha_i \beta_i}{2 \pi^2 T} \int_0^\infty \frac{e^{(E_i - \mu_i)/T} p^2 dp}{\left[e^{(E_i - \mu_i)/T} \pm 1\right]^2}
\end{equation}
For example, the baryon-strangeness susceptibility is found by setting $\alpha = B$, $\beta = S$, \begin{equation}
  \chi_{\rm BS}(T,\{\mu \}) = \sum_i \frac{g_i  B_i S_i}{2 \pi^2 T} \int_0^\infty \frac{e^{(E_i - \mu_i)/T} p^2 dp}{\left[e^{(E_i - \mu_i)/T} \pm 1\right]^2}
\end{equation}

\subsection{Boltzmann Approximation}
If quantum statistics are necessary, one needs to use the expressions
given in the previous subsection to compute relevant thermodynamic
quantities.  However, for physical hadronic masses and temperatures
below deconfinement, quantum statistics represent only a small
correction to the classical results.  The classical expressions are
advantageous due to their analytical tractability.  We will
exclusively use the Boltzmann approximation for the remainder of this
work.  For Boltzmann statistics, we can neglect the $\pm1$ in the
distribution function.  \commentout{ leading to
\begin{eqnarray}
  n_i(T,\mu_i) &=& \frac{g_i e^{\mu_i/T}}{2 \pi^2}\int_0^\infty dp p^2 e^{-E_i/T}\\
  s_i(T,\mu_i) &=& \frac{g_i e^{\mu_i/T}}{2 \pi^2 T} \int_0^\infty \left[E_i - \mu_i + \frac{p^2}{3 E_i} \right]p^2 e^{-E_i/T}dp \nonumber \\ \\
  \chi_{\alpha \beta}(T,\{\mu\}) &=& \sum_i \frac{g_i \alpha_i \beta_i e^{\mu_i/T}}{2 \pi^2 T} \int_0^\infty dp p^2 e^{-E_i/T} \\
  &=& \frac{1}{T}\sum_i \alpha_i \beta_i n_i(T,\mu_i)
\end{eqnarray}
}
The remaining integrals can be done analytically leading to
\begin{eqnarray}
  n_i(T,\mu_i) &=& n_i^{\rm FE}(T) e^{\mu_i/T} \label{nMB}\\
  s_i(T,\mu_i) &=& n_i(T,\mu_i) \left[\frac{m_i}{T}\frac{K_3(m_i/T)}{K_2(m_i/T)} - \frac{\mu_i}{T} \right]
  \label{sMB}
\end{eqnarray}
where $K_2(x)$ is a modified Bessel function.  The function
$n_i^{\rm FE}(T)$ is the number density in full equilibrium (FE), and
where all chemical potentials vanish $\mu_B = \mu_Q = \mu_S = 0$
(which is the case shortly after thermalization in a high energy heavy
ion collision).
\begin{eqnarray}
  n_i^{\rm FE}(T) &\equiv&  \frac{g_i T^3}{2 \pi^2} \left(\frac{m_i}{T}\right)^2 K_2\left(\frac{m_i}{T} \right)
  \label{nFE}
\end{eqnarray}
The total entropy in full equilibrium (and where all chemical potentials vanish) is
\begin{eqnarray}
  s^{\rm FE}(T) &=& \sum_i n^{\rm FE}_i(T) \left[\frac{m_i}{T}\frac{K_3(m_i/T)}{K_2(m_i/T)} \right].
  \label{sFE}
\end{eqnarray}
The susceptibilities are given by
\begin{equation}
  \chi_{\alpha \beta}^{\rm FE}(T) = \frac{T^2}{2 \pi^2} \sum_i \alpha_i \beta_i g_i \left(\frac{m_i}{T}\right)^2 K_2\left(\frac{m_i}{T} \right)
\label{ChiMB}
\end{equation}

Given a list of hadrons with masses $m_i$, the (full equilibrium) susceptibilities as a
function of temperature can be calculated using (\ref{ChiMB}).
Calculations of this sort were carried out and compared with lattice
QCD results in \cite{Bazavov:2012jq}.


\section{Full chemical Freeze Out (FFO)}
\label{FFOSec}
The results of the previous section are applicable for a system in
full chemical equilibrium.  The system created in a heavy ion
collision is not static and has a finite size; it does not remain in
chemical equilibrium throughout its entire evolution.  Eventually,
number changing processes freeze out, and the only remaining
interactions between hadrons are elastic scattering processes.  During
this phase, the system is chemically frozen out (but still in local
\emph{thermal} equilibrium).

A first attempt to model this phase can be made by considering the total number of
each species of hadron ($N_i \equiv n_i V$ where $V$ is the volume of
the system) to be fixed \cite{Hirano:2002ds}.  We use the superscript
``FFO'' to denote ``Full Freeze-Out'', meaning \emph{each} hadron
species has a fixed number after chemical freeze-out.  In reality,
some particles decay; we include this effect in the next section.

To maintain this chemical freeze-out condition, we must introduce
additional chemical potentials corresponding to the new conserved
particle numbers.  Assuming \emph{each} species of hadron is conserved
after chemical freeze-out, there will be one new chemical potential
for each hadron.  Hence:
\begin{eqnarray}
  \mu_i = Q_i \mu_Q + B_i \mu_B + S_i \mu_S + \mu^{\rm FFO}_{\rm i}
\end{eqnarray}
The additional chemical potential $\mu^{\rm FFO}_i$ vanishes while the
system is in chemical equilibrium, but becomes active after chemical
freeze-out.  One can think of this as suddenly associating a conserved
``charge'' with each species of hadron.  For example, $\pi^+$
particles have a (+1) $\pi^+$ ``charge'', $\pi^0$ particles have a
(+1) $\pi^0$ ``charge'' (and zero $\pi^+, \pi^-$ ``charge''), etc...

In order to remove any dependence on the volume of the system, it is
more convenient to impose the condition
\begin{equation}
  \frac{n_i(T,\mu_i^{\rm FFO})}{s(T,\{\mu^{\rm FFO}\})} = \frac{n^{\rm FE}_i(T_{\rm ch})}{s^{\rm FE}(T_{\rm ch})}
\label{CFOcondition}
\end{equation}
where the right hand side is a constant.  The temperature $T_{\rm ch}$
is the chemical freeze-out temperature below which number changing
processes are no longer effective.  This condition is equivalent to
$N_i = \rm{constant}$ because (neglecting dissipative corrections) the
total entropy $s V$ is also conserved.

\subsection{Analytical Results}
The central problem is how to determine the temperature dependence of
all of the chemical potentials $\mu^{\rm FFO}_i$ such that condition
(\ref{CFOcondition}) is satisfied. In this subsection, we give some
analytical formulas for the chemical potentials and susceptibilities
which (to the best of our knowledge) have not previously appeared in
the literature.

Assuming $\mu_B = \mu_Q = \mu_S = 0$ we have $\mu_j = \mu^{\rm FFO}_j$.
If we have a system of $N_{\rm tot}$ species of particles, there are
$N_{\rm tot}-1$ constraints of the form
\begin{eqnarray}
  \frac{n_j(T,\mu_j^{\rm FFO})}{n_i(T,\mu_i^{\rm FFO})} = \frac{n_j^{\rm FE}(T_{\rm ch})}{n_i^{\rm FE}(T_{\rm ch})}
\end{eqnarray}
Substituting (\ref{nMB}) and (\ref{nFE}), one can solve for the difference
\begin{equation}
  \frac{\mu^{\rm FFO}_j(T) - \mu^{\rm FFO}_i(T)}{T} = \ln \left[\frac{ K_2(m_i/T) K_2(m_j/T_{\rm ch})}{K_2(m_i/T_{\rm ch})K_2(m_j/T)} \right]
  \label{Xconstraint}
\end{equation}
Note that at $T = T_{\rm ch}$, the difference vanishes implying all of
the chemical potentials are the same at that point (of course, by
definition they should all vanish at $T = T_{\rm ch}$ which we will
show shortly).  

\begin{figure*}[!htpb]
  \centering
  \includegraphics[width = \textwidth, trim = 24mm 0mm 15mm 0mm, clip=true]{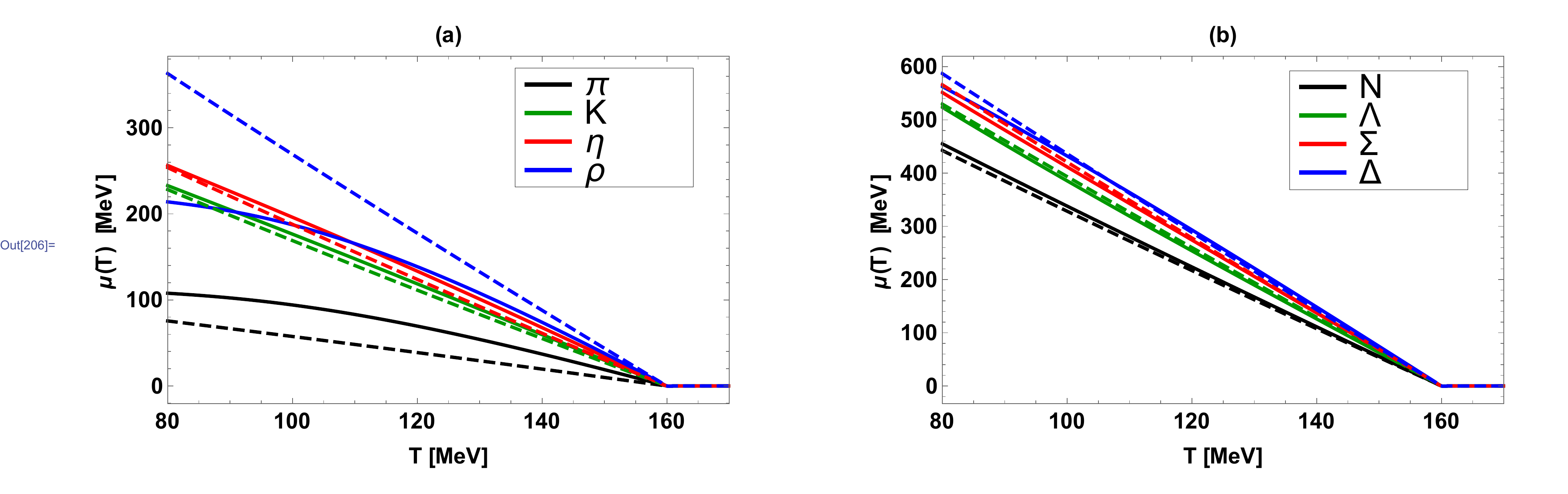}
  \caption{(Color Online) Results for the particle chemical potentials
    $\mu_i^{\rm FFO}$ and $\mu_i^{\rm PCE}$ which enforce the chemical
    freeze-out condition.  Shown are the chemical potentials for the
    four lightest mesons (a) and the four lightest baryons (b).  The
    dashed lines are the chemical potentials given the FFO freeze-out
    condition, the solid lines are the chemical potentials given the
    PCE freeze-out condition.  We assumed $T_{\rm ch} = 160$ MeV.}
  \label{MuCFO}
\end{figure*}

The final constraint is found by summing (\ref{sMB})
over all hadrons and dividing both sides by $s$, which leads to
\begin{eqnarray}
  1 = \sum_j \frac{n_j(T,\mu_j^{\rm FFO})}{s(T,\{\mu^{\rm FFO}\})}\left[\frac{m_j}{T}\frac{K_3(m_j/T)}{K_2(m_j/T)} - \frac{\mu^{\rm FFO}_j}{T} \right]
\end{eqnarray}
Substituting (\ref{CFOcondition}), the expressions for $n_j$ and $s$ in full equilibrium (\ref{nFE}),(\ref{sFE}), and the constraint (\ref{Xconstraint}) one can solve analytically for $\mu_i^{\rm FFO}$
\begin{eqnarray}
\frac{\mu_i^{\rm FFO}(T)}{T} =\ln \left[\frac{K_2(m_i/T_{\rm ch})}{K_2(m_i/T)} \right] - \Delta(T) 
\label{muFFOsolution}
\end{eqnarray}
where $\Delta(T)$ is a temperature dependent offset which is independent of the
mass $m_i$
\begin{equation}
  \Delta(T) \equiv \frac{\sum_j g_j m_j^2 K_2(m_j/T_{\rm ch})
      \left[G\left(\frac{m_j}{T}\right) - G\left(\frac{m_j}{T_{\rm ch}}\right)  \right]}
    {\sum_j g_j m_j^2 K_2(m_j/T_{\rm ch})}
\end{equation}
and we have defined
\begin{eqnarray}
  G(x) \equiv \frac{x K_3(x)}{K_2(x)} + \ln \left[K_2(x) \right]
\end{eqnarray}
Plots of the chemical potentials as a function of $T$ are shown in Fig. \ref{MuCFO} assuming $T_{\rm ch} = 160 $ MeV.

Once $T_{\rm ch}$ is specified, and a set of particles to include in
the HRG is fixed, one can use (\ref{muFFOsolution}) to
determine the chemical potential for each species of particle as a
function of $T$.  When these are known, one can go back and compute
the susceptibilities by inserting the correct $\mu^{\rm FFO}(T)$ for each
species of particle,
\begin{eqnarray}
  \chi_{\alpha \beta}^{\rm FFO}(T) = \frac{1}{T}\sum_i \alpha_i \beta_i n_i(T, \mu_i^{\rm FFO}(T))
  \label{chiFFOformula}
\end{eqnarray}
which can be written
\begin{equation}
  \chi_{\alpha \beta}^{\rm FFO}(T) = \frac{T^2e^{-\Delta(T)}}{2\pi^2}\sum_i \alpha_i \beta_i g_i 
  \left(\frac{m_i}{T}\right)^2 K_2\left(\frac{m_i}{T_{\rm ch}}\right).
\end{equation}
This analytical formula for susceptibility (and associated formulae for the 
chemical potentials) in ``full freeze-out'' is one of our main results. 

\subsection{Approximate Chemical Potentials for $m_i \gg T$}
In our calculations, we use (\ref{muFFOsolution}) to compute the
chemical potentials in the case of full freeze-out.  However, to
develop our intuition, it is useful to consider an approximation which
leads to simpler results.  The HRG model should be applicable for
temperatures of the order 100 MeV $\lesssim$ T $\lesssim$ 180 MeV,
below deconfinement and above kinetic freeze-out.  For these
temperatures, most hadrons have $m \gg T$ except for the lowest mass
ones (pions).

For $m \gg T$, one can use the asymptotic approximation for the
modified Bessel functions
\begin{eqnarray}
  K_n(x) \approx \sqrt{\frac{\pi}{2x}}e^{-x} \left[1 + \mathcal{O}\left(\frac{1}{x}\right) \right]
\end{eqnarray}
If we assume that $m_i \gg T_{\rm ch} > T$, and keep only the leading
order in the asymptotic expansion, the chemical potentials are found
to be approximately linear in $m$ with a temperature dependent offset
\begin{equation}
  \mu^{\rm FFO}_i(T) \approx m_i \left[1 - \frac{T}{T_{\rm ch}} \right] - T\left[\ln\left(\sqrt{\frac{T}{T_{\rm ch}}} \right) + \Delta (T) \right].
\end{equation}
A similar approximation was used in previous calculations
\cite{Teaney:2002aj}.  Unfortunately, a simple approximation for
$\Delta$ is more difficult to come by since the sum over $j$ necessary
to compute $\Delta$ runs over all hadrons, some of which have $m \sim
T$.

Assuming a freeze-out temperature of $T_{\rm ch} = 160$ MeV, for high
mass particles ($m \sim 1$ GeV) and low temperatures ($T \sim 100$
MeV), the approximate formula differs from the analytic result by less
than 3\%. 

In summary, for physical hadron masses and temperatures realized in a
heavy ion collision, the ``full freeze out'' chemical potentials are
approximately linear with temperature (plus corrections).  Corrections
to this approximation are due to both the contributions at $m\sim T$,
and quantum statistics.

\section{Including decays - Partial Chemical Equilibrium (PCE)}
\label{PCESec}
In reality, the number of each particle is not
precisely conserved after chemical freeze-out due to the fact that
resonances decay.  For example, the number of $\rho$ mesons is not
constant, since they will primarily decay into pions.  In order to
account for this, we use the model of partial chemical equilibrium
detailed in \cite{Bebie:1991ij} which we review here for completeness.

We consider only a subset of particles to be stable (i.e. those with
lifetimes much longer than timescale of a heavy ion collision).  We
denote number of stable particles and the total number of particles as
$N_S$, $N_{\rm tot}$ respectively.  We follow \cite{Teaney:2002aj} in choosing
the stable hadrons to be $\pi, K, \eta, \omega, N, \eta', \phi,
\Lambda, \Sigma, \Xi, \Omega$ (we include all members of any isospin multiplet and all antibaryons).  Each of these stable particles has an
associated chemical potential $\bar{\mu}^{\rm PCE}_i$.  The
generalization of (\ref{muiFFO}) is now
\begin{eqnarray}
  \mu^{\rm PCE}_i = Q_i \mu_Q + B_i \mu_B + S_i \mu_S + \sum_{j, \rm stable} d_{ij} \bar{\mu}_j^{\rm PCE}.
  \label{muPCE}
\end{eqnarray}
Note that the summation runs only over stable particles.  The
coefficients $d_{ij}$ are the ``charges'' associated with the stable
conserved chemical potentials.  For example, whereas in the full
freeze-out case, a $\rho^+$ particle would have (+1) ``$\rho^+$
charge'', in the PCE case a $\rho^+$ particle has (+1) ``$\pi^+$
charge'' and (+1) ``$\pi^0$ charge'' since a $\rho^+$ decays to $\pi^+
\pi^0$ 100\% of the time.

\begin{figure*}[!htpb]
  \centering
  \includegraphics[width=\textwidth, trim = 24mm 0mm 16mm 0mm, clip=true]{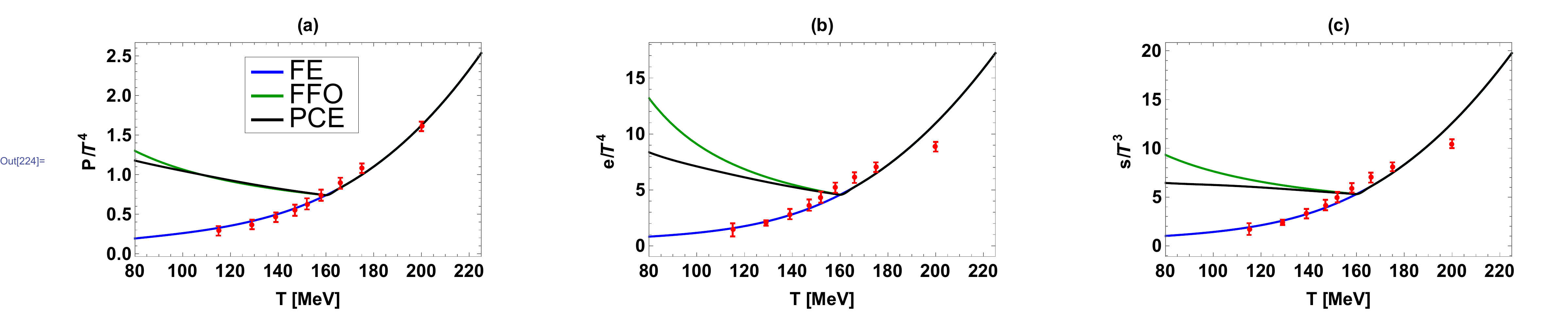}
  \caption{(color online) Dimensionless thermodynamic quantities pressure (a), energy density (b), and entropy density (c) in the case of full chemical equilibrium (FE), full freeze-out (FFO), and partial chemical equilibrium (PCE) assuming $T_{\rm ch} = 160$ MeV.  The red points are lattice data adapted from \cite{Borsanyi:2010cj}}
  \label{ePsFig}
\end{figure*}

More generally, the $d_{ij}$ coefficients are found by examining the
decay rates of particle $i$ which result in one (or more) stable
particles $j$.  Computation of the $d_{ij}$ requires branching ratios
for each hadron and resonance included in the HRG.  In Appendix
\ref{PDGappendix} we provide more details on the determination of
these decay rates, many of which are not measured precisely.  The
precise value $d_{ij}$ is found by multiplying the branching fraction
for the decay by the number of stable particles $j$ formed in that
decay, and finally summing over all decay modes of $i$.  In other
words, $d_{ij}$ is the average number of stable particles $j$ formed
from a decay of particle $i$.  A stable particle can be thought of one
which decays into itself 100\% of the time, so for example $d_{\pi^0
  i} = \delta_{\pi^0 i}$.

It is helpful to envision a matrix, $\mathbf{d}$, which has $N$ rows,
and $N_S$ columns. In appendix \ref{dijappendix} we provide a portion
of this matrix, (and hence give a subset of the $d_{ij}$
coefficients).

The quantity which is now conserved after chemical freeze-out is
\begin{eqnarray}
  \bar{N}_i = \sum_j  N_j d_{ji}
  \label{NPCE}
\end{eqnarray}
which is the ``effective number of stable particles'', including those
hidden inside the unstable particles which will eventually decay.  For example,
if we had a system of only $\rho$ and $\pi$ particles, the number
\begin{eqnarray}
  \bar{N}_{\pi^+} = N_{\pi^+} + N_{\rho^+} 
\end{eqnarray}
would be conserved after chemical freeze-out, since a $\rho^+$ effectively counts as one $\pi^+$.  Similarly, the quantity
\begin{eqnarray}
  \bar{N}_{\pi^0} =   N_{\pi^0} + N_{\rho^+} + N_{\rho^-}
\end{eqnarray}
is conserved since both $\rho^+$ and $\rho^-$ effectively count as one
$\pi^0$.

As before, to remove any dependence on the volume of the system, the freeze-out condition is
implemented using intensive quantities
\begin{eqnarray}
  \frac{\nbar_i(T,\{\bar{\mu}^{\rm PCE}\})}{s(T,\{\bar{\mu}^{\rm PCE}\})} = \frac{\nbar_i^{\rm FE}(T_{\rm ch})}{s^{\rm FE}(T_{\rm ch})} \equiv \bar{C}_i
\end{eqnarray}
This criteria is more complicated than the full freeze-out case, since
$\bar{n}_i$ now depends on multiple chemical potentials.  As before, we
set up a system of $N_S$ algebraic equations for $\{\bar{\mu}^{\rm PCE}\}$.
There are $N_S-1$ constraints which can be written without any reference to $s$:
\begin{eqnarray}
  \frac{\bar{n}_i(T,\{\bar{\mu}^{\rm PCE}\})}{n_{\Omega}(T,\bar{\mu}^{\rm PCE}_\Omega)} = \frac{\bar{C}_i}{\bar{C}_\Omega}
  \label{nbarconstraint}
\end{eqnarray}
It is advantageous to single out the heaviest stable particle
($\Omega$) because it is both stable ($d_{\Omega i} = \delta_{\Omega
  i}$), and none of the other particles we consider decay into
\emph{it} ($d_{i \Omega} = \delta_{i \Omega}$).  Hence,
$\bar{n}_{\Omega}(T,\{\bar{\mu}^{\rm PCE}\}) =
n_\Omega(T,\bar{\mu}^{\rm PCE}_\Omega)$.  To make the dependence on
the chemical potentials explicit, we can re-write these equations as
\begin{equation}
   \bar{C}_i n_{\Omega}^{\rm FE}(T) e^{\frac{\bar{\mu}_\Omega^{\rm PCE}}{T}} = \bar{C}_\Omega \sum_j n_j^{\rm FE}(T) d_{ji} \exp\left[\sum_{k}\frac{d_{jk}\bar{\mu}^{\rm PCE}_k}{T}\right] 
\label{PCEconstraint1}
\end{equation}
One more equation is necessary to close the set.  We take it to be
\begin{eqnarray}
  \bar{C}_\Omega \frac{s(T,\{\bar{\mu}^{\rm PCE}\})}{n_\Omega(T,\bar{\mu}_\Omega^{\rm PCE})} = 1
\end{eqnarray}
Substituting (\ref{sMB}), and making use of (\ref{nbarconstraint}), we find
\begin{equation}
  \sum_i  \frac{s_i^{\rm FE}(T)}{n_\Omega^{\rm FE}(T)} \exp\left[\sum_{k}\frac{d_{ik}\bar{\mu}^{\rm PCE}_k}{T}\right]  = e^{\frac{\bar{\mu}_\Omega^{\rm PCE}}{T}} \left[ 1 +  \sum_{i} \frac{\bar{C}_i \bar{\mu}_i^{\rm PCE}}{T} \right]
\label{PCEconstraint2}
\end{equation}
The right hand side of (\ref{PCEconstraint2}) makes use of the fact that
\begin{eqnarray}
  \sum_{i = 1}^{N_{\rm tot}} \mu^{\rm PCE}_i n_i = \sum_{i, \rm stable} \bar{\mu}^{\rm PCE}_i \bar{n}_i.
\end{eqnarray}
The second sum runs only over the stable particles.  This relation is straightforward to show given (\ref{muPCE}) and (\ref{NPCE}).  

Eqs. (\ref{PCEconstraint1}) and (\ref{PCEconstraint2}) form a
nonlinear system of $N_S$ algebraic equations for the set of chemical
potentials $\{\bar{\mu}^{\rm PCE}\}$.  Unlike the ``Full Freeze-Out''
case, we have been unable to solve this system analytically.  The
system can be solved numerically (for example) with a matrix
formulation of Newton's method, but in practice we find the most
efficient way to solve the system is with Mathematica's built-in
\texttt{FindRoot[]} function \cite{MMA10}.  Results for both
light mesons and baryons are shown in Fig. \ref{MuCFO}

\section{Results and discussion}
\label{ResultsSec}
With the results for the chemical potentials of each particle in hand,
we can then compute thermodynamic quantities of interest.  The energy
density, pressure and entropy density are shown in Fig. \ref{ePsFig}.  We
also include lattice data in our plots; we re-emphasize that the PCE
and FFO results are \emph{not} expected to agree with the lattice,
since lattice calculations presuppose full chemical equilibrium.

\begin{figure*}[!htpb]
  \centering
  \includegraphics[width= 0.96\textwidth, trim = 23mm 0mm 14mm 0mm, clip=true]{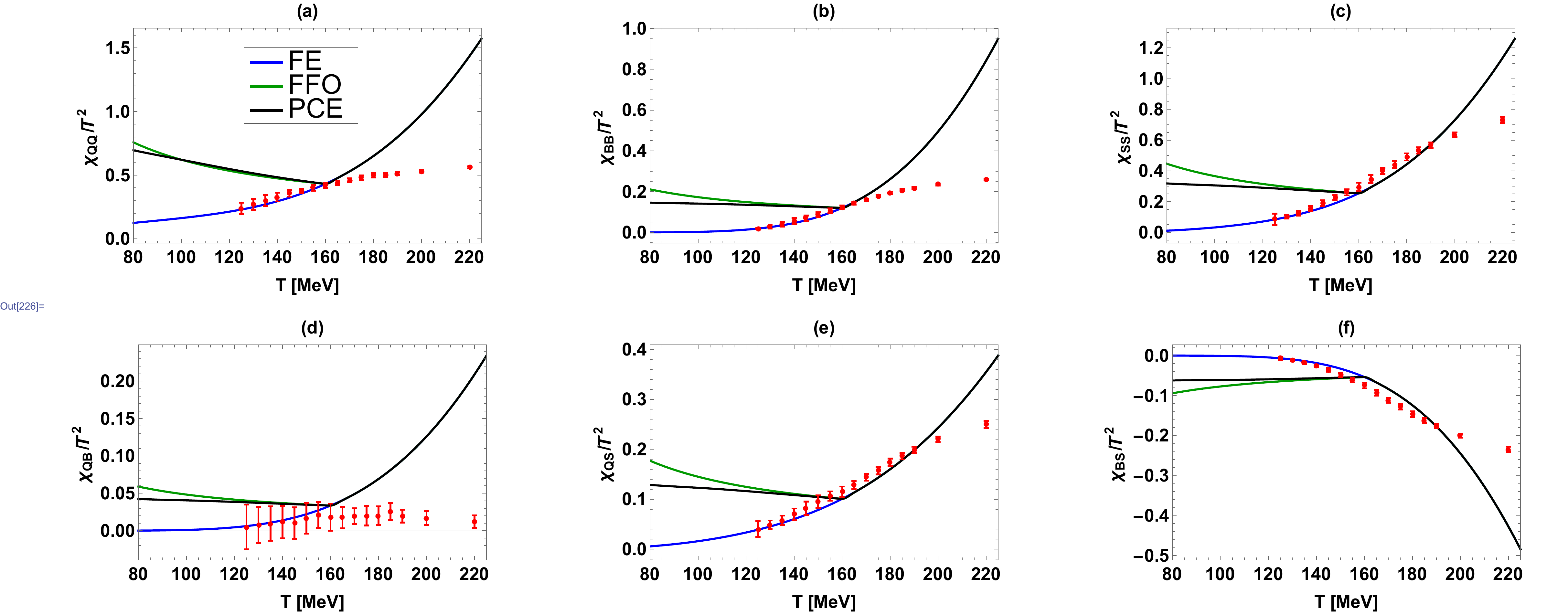}
  \caption{(color online) Dimensionless susceptibilities in the case of full chemical
    equilibrium (FE), full freeze-out (FFO), and partial chemical
    equilibrium (PCE) assuming $T_{\rm ch} = 160$ MeV.  The red points are lattice data adapted from
    \cite{Borsanyi:2011sw}.   }
  \label{ChiFO}
\end{figure*}

The susceptibilities can be seen in Fig \ref{ChiFO}.  Note that the
inclusion of chemical freeze-out can have a large effect compared to
a fully equilibrated system, especially at low temperatures where the full
equilibrium susceptibilities are approximately vanishing.  With regard
to the inclusion of decays via PCE, we see that with only a few
exceptions, the PCE scenario amounts to roughly an $\mathcal{O}$(10\%)
correction to the FFO scenario, as long as $T > 100 $ MeV.

\begin{figure*}[!htpb]
  \centering
  \includegraphics[width=0.96\textwidth, trim = 22mm 0mm 11mm 0mm, clip=true]{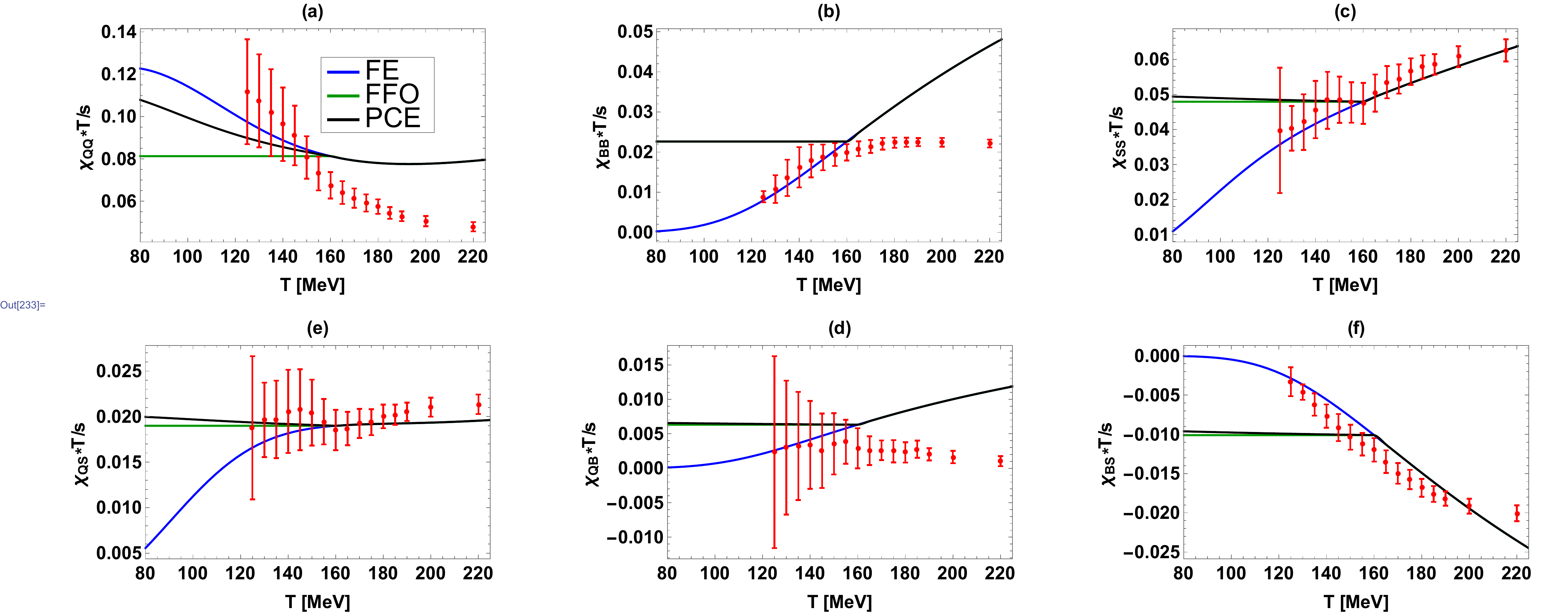}
  \caption{(color online) Susceptibility to entropy density ratio in
    the case of full chemical equilibrium (FE), full freeze-out (FFO),
    and partial chemical equilibrium (PCE) assuming $T_{\rm ch} = 160$ MeV.  The red points are
    lattice data adapted from \cite{Borsanyi:2011sw}. }
  \label{ChioverSFO}
\end{figure*}

When considering the magnitude of hydrodynamic fluctuations, the
quantity $\chi T/s$ is the relevant one \cite{Ling:2013ksb}.  We plot
this combination in Fig. \ref{ChioverSFO}.
At temperatures below $T_{\rm ch}$, $\chi^{\rm FFO}_{\alpha \beta} T /
s$ is exactly constant (since $\chi \sim \sum_i n_i$ and $n_i/s$ is
constant).  The inclusion of partial chemical equilibrium amounts to a
very small correction in most cases.

The baryon-baryon susceptibility is especially interesting.  In the
case of $\chi_{BB}/s$, the PCE and FFO curves lie on top of each
other, and (within our numerical precision) the numerical values are
exactly the same.  We suspect that $(\chi_{BB}/s)^{\rm CFO} =
(\chi_{BB}/s)^{\rm PCE}$ but were unable to show this
analytically.  It is also interesting to note that if we switch from
the HRG model to the lattice data at around 160 MeV, then $\chi_{BB} T
/ s$ is approximately constant throughout the entire range $80\,{\rm
  MeV} \leq T \leq 220$ MeV, which is simple to deal with
analytically.

The BB susceptibility is also phenomenologically interesting, as it
is important for net proton fluctuations which are used to search for
the critical point.  It is quite tempting to determine whether the
inclusion of chemical freeze-out can improve the agreement of the HRG
with the STAR measurements of these fluctuations
\cite{Fukushima:2014lfa, Albright:2015uua}.  Simple
considerations show that for any additional chemical potentials to
have an effect on the skewness or kurtosis, one must move beyond the
Boltzmann approximation used in this paper.  In our notation, the skewness 
measured in experiment is
\begin{eqnarray}
  S \sigma \equiv \frac{T}{\chi_{BB}} \pd{\chi_{BB}}{\mu_B}. 
\end{eqnarray}
Using (\ref{chiFFOformula}), and noting that all baryons have $B_i^2 = 1$, 
\begin{eqnarray}
  \chi_{BB}T &=& \sum_i n_i(T, \mu_i) \\
  \pd{\chi_{BB}}{\mu_B} &=& \frac{1}{T^2}\sum_i B_i n_i(T, \mu_i)
\end{eqnarray}
Breaking the sum over $i$ into baryons and anti-baryons and noting the only 
difference in the two is the sign of $\mu_B$ and $B_i$, we have
\begin{eqnarray}
  \pd{\chi_{BB}}{\mu_B} &=& \frac{2}{T^2} \sinh(\mu_B/T) \sum_{i,\rm baryons} n_i(T,\mu_i)\Bigl.\Bigr|_{\mu_B = 0}\\
  \chi_{BB}T &=& 2 \cosh(\mu_B/T) \sum_{\rm i,baryons} n_i(T,\mu_i)\Bigl. \Bigr|_{\mu_B = 0}.
\end{eqnarray}
And hence
\begin{eqnarray}
  S\sigma = \tanh(\mu_B/T) 
\end{eqnarray}
which is independent of whether the system is in chemical equilibrium
or frozen out.  Similarly, the kurtosis $\kappa \sigma^2$ can be shown
to be equal to 1 whether or not one employs chemical freeze-out.
Thus, whether or not chemical freeze-out of hadrons can improve the
agreement of the HRG with STAR measurements requires more careful
consideration which is beyond the scope of this paper; we defer it to future work.
\vspace{4mm}
\section{Conclusion}
\label{ConclusionSec}
We have computed all six components of the susceptibility matrix for
three conserved charges: electric charge, baryon number, and
strangeness.  We have done so using the hadron-resonance gas model
employing two different implementations of chemical freeze-out. 

In almost all cases, the inclusion of chemical freeze-out enhances $\chi_{\alpha
  \beta} T /s$ at low temperatures relative to the full
equilibrium case.  Thus we expect that the fluctuations at low
temperatures will generally be larger than those found if one
exclusively uses thermodynamic data from the lattice.  This appears 
consistent with the findings of \cite{Albright:2015uua}.  The
exception is the diagonal electric charge susceptibility $\chi_{QQ}
T/s$ which is a decreasing function of $T$.  Thus, the inclusion of
chemical freeze-out will tend to \emph{reduce} electric charge
fluctuations relative to equilibrium.

With regard to correlations and balance functions measured in
experiment \cite{Abelev:2013csa, Aggarwal:2010ya}: as compared to full
equilibrium, the inclusion of chemical freeze-out will tend to narrow
the pseudorapidity ($\Delta \eta$) width of the balance function (for
a given diffusion constant).  The reason is that enhanced fluctuations
at late times (low temperatures) have less time to diffuse.  The
inclusion of resonance decays via PCE is unlikely to affect the
balance functions \emph{except} in the case of electric charge
fluctuations.  As seen in Fig. \ref{ChioverSFO} (a), for
$\chi_{QQ}T/s$, partial chemical equilibrium tends to significantly
\emph{reduce} the effect of chemical freeze-out, as the PCE curve lies
closer to the full equilibrium curve than the one for full freeze-out.

In summary, the inclusion of chemical freeze-out is an important difference
between the thermodynamics of a physical heavy ion collision and
lattice QCD calculations.  This effect has implications for (among
other things) particle correlations and balance functions, and the
search for the QCD critical point.  Work to explore these implications
is already underway.

\section*{Acknowledgments}
We thank K. Dusling and C. Young for helpful discussions.  We are
especially grateful to M. Albright and P. Huovinen for sharing
resources related to particles included in the hadron-resonance gas
model.

  \bibliography{Bibliography/PCESusceptibility}

\providecommand{\href}[2]{#2}\begingroup\raggedright\begin{thebibliography}{10}

\bibitem{Gale:2013da}
C.~Gale, S.~Jeon, and B.~Schenke, ``{Hydrodynamic Modeling of Heavy-Ion
  Collisions},'' \href{http://dx.doi.org/10.1142/S0217751X13400113}{{\em
  Int.J.Mod.Phys.} {\bf A28} (2013)  1340011},
\href{http://arxiv.org/abs/1301.5893}{{\tt arXiv:1301.5893 [nucl-th]}}.

\bibitem{Stephanov:1998dy}
M.~A. Stephanov, K.~Rajagopal, and E.~V. Shuryak, ``{Signatures of the
  tricritical point in QCD},''
  \href{http://dx.doi.org/10.1103/PhysRevLett.81.4816}{{\em Phys. Rev. Lett.}
  {\bf 81} (1998)  4816--4819},
\href{http://arxiv.org/abs/hep-ph/9806219}{{\tt arXiv:hep-ph/9806219
  [hep-ph]}}.

\bibitem{Stephanov:1999zu}
M.~A. Stephanov, K.~Rajagopal, and E.~V. Shuryak, ``{Event-by-event
  fluctuations in heavy ion collisions and the QCD critical point},''
  \href{http://dx.doi.org/10.1103/PhysRevD.60.114028}{{\em Phys. Rev.} {\bf
  D60} (1999)  114028},
\href{http://arxiv.org/abs/hep-ph/9903292}{{\tt arXiv:hep-ph/9903292
  [hep-ph]}}.

\bibitem{Hatta:2003wn}
Y.~Hatta and M.~A. Stephanov, ``{Proton number fluctuation as a signal of the
  QCD critical endpoint},''
  \href{http://dx.doi.org/10.1103/PhysRevLett.91.102003}{{\em Phys. Rev. Lett.}
  {\bf 91} (2003)  102003}, \href{http://arxiv.org/abs/hep-ph/0302002}{{\tt
  arXiv:hep-ph/0302002 [hep-ph]}}.
[Erratum: Phys. Rev. Lett.91,129901(2003)].

\bibitem{Adamczyk:2013dal}
{\bf STAR} Collaboration, L.~Adamczyk {\em et al.}, ``{Energy Dependence of
  Moments of Net-proton Multiplicity Distributions at RHIC},''
  \href{http://dx.doi.org/10.1103/PhysRevLett.112.032302}{{\em Phys. Rev.
  Lett.} {\bf 112} (2014)  032302},
\href{http://arxiv.org/abs/1309.5681}{{\tt arXiv:1309.5681 [nucl-ex]}}.

\bibitem{Kapusta:2011gt}
J.~I. Kapusta, B.~Muller, and M.~Stephanov, ``{Relativistic Theory of
  Hydrodynamic Fluctuations with Applications to Heavy Ion Collisions},''
  \href{http://dx.doi.org/10.1103/PhysRevC.85.054906}{{\em Phys. Rev.} {\bf
  C85} (2012)  054906},
\href{http://arxiv.org/abs/1112.6405}{{\tt arXiv:1112.6405 [nucl-th]}}.

\bibitem{Springer:2012iz}
T.~Springer and M.~Stephanov, ``{Hydrodynamic fluctuations and two-point
  correlations},''
  \href{http://dx.doi.org/10.1016/j.nuclphysa.2013.02.190}{{\em Nucl. Phys.}
  {\bf A904-905} (2013)  1027c--1030c},
\href{http://arxiv.org/abs/1210.5179}{{\tt arXiv:1210.5179 [nucl-th]}}.

\bibitem{Ling:2013ksb}
B.~Ling, T.~Springer, and M.~Stephanov, ``{Hydrodynamics of charge fluctuations
  and balance functions},''
  \href{http://dx.doi.org/10.1103/PhysRevC.89.064901}{{\em Phys.Rev.} {\bf C89}
  (2014) no.~6, 064901},
\href{http://arxiv.org/abs/1310.6036}{{\tt arXiv:1310.6036 [nucl-th]}}.

\bibitem{Abelev:2013csa}
{\bf ALICE Collaboration} Collaboration, B.~Abelev {\em et al.}, ``{Charge
  correlations using the balance function in Pb-Pb collisions at
  $\sqrt{s_{NN}}$ = 2.76 TeV},''
  \href{http://dx.doi.org/10.1016/j.physletb.2013.05.039}{{\em Phys.Lett.} {\bf
  B723} (2013)  267--279},
\href{http://arxiv.org/abs/1301.3756}{{\tt arXiv:1301.3756 [nucl-ex]}}.

\bibitem{Aggarwal:2010ya}
{\bf STAR} Collaboration, M.~Aggarwal {\em et al.}, ``{Balance Functions from
  Au$+$Au, $d+$Au, and $p+p$ Collisions at $\sqrt{s_{NN}}$ = 200 GeV},''
  \href{http://dx.doi.org/10.1103/PhysRevC.82.024905}{{\em Phys.Rev.} {\bf C82}
  (2010)  024905},
\href{http://arxiv.org/abs/1005.2307}{{\tt arXiv:1005.2307 [nucl-ex]}}.

\bibitem{Borsanyi:2010cj}
S.~Borsanyi, G.~Endrodi, Z.~Fodor, A.~Jakovac, S.~D. Katz, S.~Krieg, C.~Ratti,
  and K.~K. Szabo, ``{The QCD equation of state with dynamical quarks},''
  \href{http://dx.doi.org/10.1007/JHEP11(2010)077}{{\em JHEP} {\bf 11} (2010)
  077},
\href{http://arxiv.org/abs/1007.2580}{{\tt arXiv:1007.2580 [hep-lat]}}.

\bibitem{Borsanyi:2011sw}
S.~Borsanyi, Z.~Fodor, S.~D. Katz, S.~Krieg, C.~Ratti, and K.~Szabo,
  ``{Fluctuations of conserved charges at finite temperature from lattice
  QCD},'' \href{http://dx.doi.org/10.1007/JHEP01(2012)138}{{\em JHEP} {\bf 01}
  (2012)  138},
\href{http://arxiv.org/abs/1112.4416}{{\tt arXiv:1112.4416 [hep-lat]}}.

\bibitem{Bazavov:2012jq}
{\bf HotQCD} Collaboration, A.~Bazavov {\em et al.}, ``{Fluctuations and
  Correlations of net baryon number, electric charge, and strangeness: A
  comparison of lattice QCD results with the hadron resonance gas model},''
  \href{http://dx.doi.org/10.1103/PhysRevD.86.034509}{{\em Phys.Rev.} {\bf D86}
  (2012)  034509},
\href{http://arxiv.org/abs/1203.0784}{{\tt arXiv:1203.0784 [hep-lat]}}.

\bibitem{Bebie:1991ij}
H.~Bebie, P.~Gerber, J.~Goity, and H.~Leutwyler, ``{The Role of the entropy in
  an expanding hadronic gas},''
\href{http://dx.doi.org/10.1016/0550-3213(92)90005-V}{{\em Nucl.Phys.} {\bf
  B378} (1992)  95--130}.

\bibitem{Hirano:2002ds}
T.~Hirano and K.~Tsuda, ``{Collective flow and two pion correlations from a
  relativistic hydrodynamic model with early chemical freezeout},''
  \href{http://dx.doi.org/10.1103/PhysRevC.66.054905}{{\em Phys.Rev.} {\bf C66}
  (2002)  054905},
\href{http://arxiv.org/abs/nucl-th/0205043}{{\tt arXiv:nucl-th/0205043
  [nucl-th]}}.

\bibitem{Teaney:2002aj}
D.~Teaney, ``{Chemical freezeout in heavy ion collisions},''
\href{http://arxiv.org/abs/nucl-th/0204023}{{\tt arXiv:nucl-th/0204023
  [nucl-th]}}.

\bibitem{Huovinen:2007xh}
P.~Huovinen, ``{Chemical freeze-out temperature in hydrodynamical description
  of Au+Au collisions at s(NN)**(1/2) = 200-GeV},''
  \href{http://dx.doi.org/10.1140/epja/i2007-10611-3}{{\em Eur.Phys.J.} {\bf
  A37} (2008)  121--128},
\href{http://arxiv.org/abs/0710.4379}{{\tt arXiv:0710.4379 [nucl-th]}}.

\bibitem{Huovinen:2009yb}
P.~Huovinen and P.~Petreczky, ``{QCD Equation of State and Hadron Resonance
  Gas},'' \href{http://dx.doi.org/10.1016/j.nuclphysa.2010.02.015}{{\em Nucl.
  Phys.} {\bf A837} (2010)  26--53},
\href{http://arxiv.org/abs/0912.2541}{{\tt arXiv:0912.2541 [hep-ph]}}.

\bibitem{Dashen:1969ep}
R.~Dashen, S.-K. Ma, and H.~J. Bernstein, ``{S Matrix formulation of
  statistical mechanics},''
\href{http://dx.doi.org/10.1103/PhysRev.187.345}{{\em Phys. Rev.} {\bf 187}
  (1969)  345--370}.

\bibitem{Venugopalan:1992hy}
R.~Venugopalan and M.~Prakash, ``{Thermal properties of interacting hadrons},''
\href{http://dx.doi.org/10.1016/0375-9474(92)90005-5}{{\em Nucl. Phys.} {\bf
  A546} (1992)  718--760}.

\bibitem{Becattini:2004td}
F.~Becattini, ``{What is the meaning of the statistical hadronization
  model?},'' \href{http://dx.doi.org/10.1088/1742-6596/5/1/015}{{\em J. Phys.
  Conf. Ser.} {\bf 5} (2005)  175--188},
  \href{http://arxiv.org/abs/hep-ph/0410403}{{\tt arXiv:hep-ph/0410403
  [hep-ph]}}.
[,175(2004)].

\bibitem{Agashe:2014kda}
{\bf Particle Data Group} Collaboration, K.~A. Olive {\em et al.}, ``{Review of
  Particle Physics},''
\href{http://dx.doi.org/10.1088/1674-1137/38/9/090001}{{\em Chin. Phys.} {\bf
  C38} (2014)  090001}.

\bibitem{MMA10}
Wolfram{\ }Research{\ }Inc., {\em Mathematica, v10.1}.
\newblock Champaign, IL, (2015).

\bibitem{Fukushima:2014lfa}
K.~Fukushima, ``{Hadron resonance gas and mean-field nuclear matter for baryon
  number fluctuations},''
  \href{http://dx.doi.org/10.1103/PhysRevC.91.044910}{{\em Phys. Rev.} {\bf
  C91} (2015) no.~4, 044910},
\href{http://arxiv.org/abs/1409.0698}{{\tt arXiv:1409.0698 [hep-ph]}}.

\bibitem{Albright:2015uua}
M.~Albright, J.~Kapusta, and C.~Young, ``{Baryon Number Fluctuations from a
  Crossover Equation of State Compared to Heavy-Ion Collision Measurements in
  the Beam Energy Range $\sqrt{s_{NN}}$ = 7.7 to 200 GeV},''
\href{http://arxiv.org/abs/1506.03408}{{\tt arXiv:1506.03408 [nucl-th]}}.

\bibitem{Broniowski:2015oha}
W.~Broniowski, F.~Giacosa, and V.~Begun, ``{Why the sigma meson should not be
  included in thermal models},''
\href{http://arxiv.org/abs/1506.01260}{{\tt arXiv:1506.01260 [nucl-th]}}.

\bibitem{Huovinen:PC}
P.~Huovinen. (private communication).

\bibitem{PasiOnline}
P.~Huovinen, ``Parametrization of the equation of state.''
  \url{https://wiki.bnl.gov/TECHQM/index.php/QCD_Equation_of_State}.

\bibitem{Li:2008xy}
D.-M. Li and S.~Zhou, ``{On the nature of the pi(2)(1880)},''
  \href{http://dx.doi.org/10.1103/PhysRevD.79.014014}{{\em Phys. Rev.} {\bf
  D79} (2009)  014014},
\href{http://arxiv.org/abs/0811.0918}{{\tt arXiv:0811.0918 [hep-ph]}}.

\bibitem{Pang:2015eha}
C.-Q. Pang, B.~Wang, X.~Liu, and T.~Matsuki, ``{High-spin mesons below 3
  GeV},'' \href{http://dx.doi.org/10.1103/PhysRevD.92.014012}{{\em Phys. Rev.}
  {\bf D92} (2015) no.~1, 014012},
\href{http://arxiv.org/abs/1505.04105}{{\tt arXiv:1505.04105 [hep-ph]}}.

\end{thebibliography}\endgroup

\appendix


\section{Particles, Resonances, and Decays}
\label{PDGappendix}
We include all hadrons and resonances listed in the 2014 PDG (with
rating of *** or ****) with masses $\lesssim$ 2 GeV.  The most massive
particle we consider is the $f_2(2010)$, but we omit $f_0(600)$
($\sigma$ meson) \cite{Broniowski:2015oha}. The masses and quantum
numbers of all included particles match those given in the 2014 Review
of Particle Physics provided by the PDG \cite{Agashe:2014kda}.

The measured decay modes and branching fractions of the higher mass
resonances are very uncertain.  Often decay channels are simply
labeled as ``seen'' and/or the measured branching fractions do not sum
to one.  Hence, theoretical input is required.  We rely almost exclusively on
previous work by Josef Sollfrank and Pasi Huovinen who created a table
of decays from the 2005 PDG data, supplemented by educated guesses
based on the behavior of similar resonances and the fact that the
branching ratios must sum to one \cite{Huovinen:PC, PasiOnline}.  

We have not attempted to update all branching fractions to incorporate
more recent PDG data, as our results are most sensitive to lower mass
hadrons/resonances, which were already measured precisely in 2005.
Nevertheless, we have added three new resonances which were
not present in the 2005 Review of Particle Physics: $\pi_2(1880)$, $N(1875)$ and $N(1900)$.
The branching ratios for each of these resonances are very poorly
known, hence (as in the original table) we rely on theoretical input
and educated guesses to estimate decay rates of these particles.
These branching fractions do not significantly affect our results, but
we include them here for completeness.  Our assumptions are given in Table
\ref{BrTable}, and we explain our rationale as follows.

For the $\pi_2(1880)$, no decay modes are listed in the PDG.  Our
assumed branching fractions rely on the nearby resonance $\pi_2(1670)$
and the results of the model of \cite{Li:2008xy}.  For the $N(1875)$,
the PDG listings of the measured branching fractions do not sum to
one.  We attempt consistency with the measured values, forcing the sum
to be one.  For the $N(1900)$, the measured branching fractions sum to
less than 50\%.  We assume that the remainder is in the form of decays to $N
\rho$, as is the case for the nearby resonance $N(1720)$ (which has identical quantum numbers).  Finally,
for the $a_4(2040)$, the PDG lists only a few decay modes as ``seen''.
Our input for this resonance is almost entirely theoretical, we start
from the calculations of \cite{Pang:2015eha} and then include the
$K\bar{K}$ and $\eta' \pi$ modes to make the sum of the branching
ratios one.
\renewcommand*{\arraystretch}{1.2}
\LTcapwidth=0.8 \columnwidth
\begin{longtable}[h]{ll}
    \hline
    $\pi_2(1880) \longrightarrow \rho \pi$ & (33\%)  \\*
    $\pi_2(1880) \longrightarrow f_2(1270) \pi$ & (17\%)  \\*
    $\pi_2(1880) \longrightarrow \rho \omega$ & (17\%)  \\*
    $\pi_2(1880) \longrightarrow KK^*$ & (11\%)  \\*
    $\pi_2(1880) \longrightarrow \rho(1450) \pi*$ & (11\%)  \\*
    $\pi_2(1880) \longrightarrow a(1320) \eta$ & (11\%)  \\
    \hline
    $N(1875) \longrightarrow N \pi$ & (10 \%) \\*
    $N(1875) \longrightarrow N \omega$ & (20 \%) \\*
    $N(1875) \longrightarrow \Delta \pi$ & (50 \%) \\*
    $N(1875) \longrightarrow N \pi \pi$ & (15 \%) \\*
    $N(1875) \longrightarrow N \rho $ & (5 \%) \\*
    \hline
    $N(1900) \longrightarrow N \pi$ & (6 \%) \\*
    $N(1900) \longrightarrow \rho \eta$ & (12 \%) \\*
    $N(1900) \longrightarrow \rho \omega$ & (12 \%) \\*
    $N(1900) \longrightarrow \Lambda K $ & (4 \%) \\*
    $N(1900) \longrightarrow \Sigma K $ & (6 \%) \\*
    $N(1900) \longrightarrow N \rho $ & (60 \%) \\
    \hline
    $a_4(2040) \longrightarrow \rho \omega$ &  (37\%)\\*
    $a_4(2040) \longrightarrow \rho \pi$ & (20 \%) \\*
    $a_4(2040) \longrightarrow b_2(1235) \pi$ & (22 \%) \\*
    $a_4(2040) \longrightarrow f_2(1270) \pi$ & (11 \%) \\*
    $a_4(2040) \longrightarrow K \bar{K}$ & (5 \%) \\*
    $a_4(2040) \longrightarrow \eta' \pi$ & (5 \%) \\
    \hline
    \caption{Assumed branching fractions for newly discovered resonances 
    \label{BrTable}}
  \end{longtable}

\section{Sample Decay Coefficients for Partial Chemical Equilibrium}
\label{dijappendix}
As a convenience to the reader, Table \ref{dijtable} is a sample of
the ``effective number of stable particles'' formed from each unstable
particle (i.e. some of the $d_{ij}$ coefficients).  Note that this is
only a partial list.

While we have tabulated the $d_{ij}$ coefficients for photons produced
in decays, we omit the photon when computing thermodynamic quantities
like $n$, $s$ and $\chi_{\alpha \beta}$ as it is not expected that
these photons will thermalize and they are not considered part of the
HRG model.

\begin{center}
\renewcommand*{\arraystretch}{1.2}
\LTcapwidth=\textwidth
\begin{longtable*}[H]{|c||c|c|c|c|c|c|c|c|c|c|c|c|c|c|c|c|c|c|}
  \hline
 & $\gamma$ & $\pi^+$ & $\pi^0$ & $\pi^-$ & $K^+$ & $K^-$ & $K^0$ & $\bar{K}^0$ & $\eta$ & $\omega$ & $p$ & $\bar{p}$ & $n$ & $\bar{n}$ & $\eta'$ & $\phi$ & $\Lambda$ & $\bar{\Lambda}$ 
\\
\hline
\hline
 $\rho^+$ & 0 & 1.000 & 1.000 & 0 & 0 & 0 & 0 & 0 & 0 & 0 & 0 & 0 & 0 & 0 & 0 & 0 & 0 & 0 
 \\ \hline
 $\rho^0$ & 0 & 1.000 & 0 & 1.000 & 0 & 0 & 0 & 0 & 0 & 0 & 0 & 0 & 0 & 0 & 0 & 0 & 0 & 0 
 \\ \hline
 $\rho^-$ & 0 & 0 & 1.000 & 1.000 & 0 & 0 & 0 & 0 & 0 & 0 & 0 & 0 & 0 & 0 & 0 & 0 & 0 & 0 
 \\ \hline
 $K^*(892)^+$ & 0 & 0.667 & 0.333 & 0 & 0.333 & 0 & 0.667 & 0 & 0 & 0 & 0 & 0 & 0 & 0 & 0 & 0 & 0 & 0 
 \\ \hline
 $K^*(892)^-$ & 0 & 0 & 0.333 & 0.667 & 0 & 0.333 & 0 & 0.667 & 0 & 0 & 0 & 0 & 0 & 0 & 0 & 0 & 0 & 0 
 \\ \hline
 $K^*(892)^0$ & 0 & 0 & 0.333 & 0.667 & 0.667 & 0 & 0.333 & 0 & 0 & 0 & 0 & 0 & 0 & 0 & 0 & 0 & 0 & 0 
 \\ \hline
 $\bar{K}^*(892)^0$ & 0 & 0.667 & 0.333 & 0 & 0 & 0.667 & 0 & 0.333 & 0 & 0 & 0 & 0 & 0 & 0 & 0 & 0 & 0 & 0 
 \\ \hline
 $f_0(980)$ & 0 & 0.52 & 0.52 & 0.52 & 0.11 & 0.11 & 0.11 & 0.11 & 0 & 0 & 0 & 0 & 0 & 0 & 0 & 0 & 0 & 0 
 \\ \hline
 $a_0(980)^+$ & 0 & 0.844 & 0 & 0 & 0.156 & 0 & 0 & 0.156 & 0.844 & 0 & 0 & 0 & 0 & 0 & 0 & 0 & 0 & 0 
 \\ \hline
 $a_0(980)^0$ & 0 & 0 & 0.844 & 0 & 0.078 & 0.078 & 0.078 & 0.078 & 0.844 & 0 & 0 & 0 & 0 & 0 & 0 & 0 & 0 & 0 
 \\ \hline
 $a_0(980)^-$ & 0 & 0 & 0 & 0.844 & 0 & 0.156 & 0.156 & 0 & 0.844 & 0 & 0 & 0 & 0 & 0 & 0 & 0 & 0 & 0 
 \\ \hline
 $h_1(1170)$ & 0 & 1.000 & 1.000 & 1.000 & 0 & 0 & 0 & 0 & 0 & 0 & 0 & 0 & 0 & 0 & 0 & 0 & 0 & 0 
 \\ \hline
 $b_1(1235)^+$ & 0 & 1.000 & 0 & 0 & 0 & 0 & 0 & 0 & 0 & 1.000 & 0 & 0 & 0 & 0 & 0 & 0 & 0 & 0 
 \\ \hline
 $b_1(1235)^0$ & 0 & 0 & 1.000 & 0 & 0 & 0 & 0 & 0 & 0 & 1.000 & 0 & 0 & 0 & 0 & 0 & 0 & 0 & 0 
 \\ \hline
 $b_1(1235)^-$ & 0 & 0 & 0 & 1.000 & 0 & 0 & 0 & 0 & 0 & 1.000 & 0 & 0 & 0 & 0 & 0 & 0 & 0 & 0 
 \\ \hline
 $a_1(1260)^+$ & 0 & 1.000.5 & 1.000 & 0.5 & 0 & 0 & 0 & 0 & 0 & 0 & 0 & 0 & 0 & 0 & 0 & 0 & 0 & 0 
 \\ \hline
 $a_1(1260)^0$ & 0 & 1.000 & 1.000 & 1.000 & 0 & 0 & 0 & 0 & 0 & 0 & 0 & 0 & 0 & 0 & 0 & 0 & 0 & 0 
 \\ \hline
 $a_1(1260)^-$ & 0 & 0.5 & 1.000 & 1.5 & 0 & 0 & 0 & 0 & 0 & 0 & 0 & 0 & 0 & 0 & 0 & 0 & 0 & 0 
 \\ \hline
 $\Delta^{++}$ & 0 & 1.000 & 0 & 0 & 0 & 0 & 0 & 0 & 0 & 0 & 1.000 & 0 & 0 & 0 & 0 & 0 & 0 & 0 
 \\ \hline
 $\bar{\Delta}^{++}$ & 0 & 0 & 0 & 1.000 & 0 & 0 & 0 & 0 & 0 & 0 & 0 & 1.000 & 0 & 0 & 0 & 0 & 0 & 0 
 \\ \hline
 $\Delta^{+}$ & 0.006 & 0.331 & 0.663 & 0 & 0 & 0 & 0 & 0 & 0 & 0 & 0.669 & 0 & 0.331 & 0 & 0 & 0 & 0 & 0 
 \\ \hline
 $\bar{\Delta}^{+}$ & 0.006 & 0 & 0.663 & 0.331 & 0 & 0 & 0 & 0 & 0 & 0 & 0 & 0.669 & 0 & 0.331 & 0 & 0 & 0 & 0 
 \\ \hline
 $\Delta^0$ & 0.006 & 0 & 0.663 & 0.331 & 0 & 0 & 0 & 0 & 0 & 0 & 0.331 & 0 & 0.669 & 0 & 0 & 0 & 0 & 0 
 \\ \hline
 $\bar{\Delta}^0$ & 0.006 & 0.331 & 0.663 & 0 & 0 & 0 & 0 & 0 & 0 & 0 & 0 & 0.331 & 0 & 0.669 & 0 & 0 & 0 & 0 
 \\ \hline
 $\Delta^-$ & 0 & 0 & 0 & 1.000 & 0 & 0 & 0 & 0 & 0 & 0 & 0 & 0 & 1.000 & 0 & 0 & 0 & 0 & 0 
 \\ \hline
 $\bar{\Delta}^-$ & 0 & 1.000 & 0 & 0 & 0 & 0 & 0 & 0 & 0 & 0 & 0 & 0 & 0 & 1.000 & 0 & 0 & 0 & 0 
 \\ \hline
 \caption{Incomplete list of the coefficients $d_{ij}$.  The top row
   lists the 18 lightest stable particles.  The first column lists
   the lightest 26 unstable particles.   The table entry of the $ith$
   row, $jth$ column is the coefficient $d_{ij}$.  We omit columns for the
   heavier $\Sigma, \Xi, \Omega$ (their entries are all zero for the
   selection of unstable particles listed).  
\label{dijtable}
}
\end{longtable*}
\end{center}


\end{document}